# Enhancement of interfacial thermal conductance of SiC by overlapped carbon nanotubes and intertube atoms

Chengcheng Deng[1, #], Xiaoxiang Yu[1, #], Xiaoming Huang[1, *], Nuo Yang[2, 3, *]

1 School of Energy and Power Engineering, Huazhong University of Science and Technology, Wuhan 430074, P. R. China

2 State Key Laboratory of Coal Combustion, Huazhong University of Science and Technology, Wuhan 430074, P. R. China

3 Nano Interface Center for Energy (NICE), School of Energy and Power Engineering, Huazhong University of Science and Technology, Wuhan 430074, P. R. China

# C. D. and X. Y. contributed equally to this work.

* To whom correspondence should be addressed. E-mail: nuo@hust.edu.cn (NY), xmhuang@hust.edu.cn (XH)

# Abstract

We proposed a new way, adding intertube atoms, to enhance interfacial thermal conductance (ITC) between SiC-carbon nanotube (CNT) array structure. Non-equilibrium molecular dynamics method was used to study the ITC. The results show that the intertube atoms can significantly enhance the ITC. The dependence of ITC on both the temperature and the number of intertube atoms are shown. The mechanism is analyzed by calculating probability distributions of atomic forces and vibrational density of states. Our study may provide some guidance on enhancing the ITC of CNT-based composites.

**Introduction**

With the extensions of human's exploration and application areas, some indispensable materials are demanded to overcome hostile service environments in special applications such as high-temperature and irradiation environments. The silicon carbide (SiC) is emerging as a potential high-performance material which can be applied in electronics, nuclear reactors, aerospace et al.,[1-5] because of its wide band gap, large breakdown field, good mechanical property, and high thermal conductivity.[1,6,7] In real applications, it is especially required to estimate the heat dissipation in SiC devices and composites, where the thermal interface resistance dominates in phonon propagating. Recently, the studies on thermal transport through the interface, which exists broadly in electronics and composites, have received more attentions.[8-18]

Carbon nanotubes (CNTs) have been theoretically predicted and experimentally proved to have ultrahigh thermal conductivity along the axial direction.[19-25] Therefore, CNTs are currently investigated for use as thermal interface materials both theoretically and experimentally.[26,27] It was simulated by molecular dynamics (MD) that the CNT arrays are directly bonded as a bridge between two solid surfaces.[28] However, this structure is hardly realized because CNTs can not grow on both surfaces at the same time. In the experimental study, CNT arrays were respectively grown on each SiC surface and then formed a composite sandwich structure, which has a low thermal conductivity as 0.5 W/m-K.[29] Although a single CNT has a good thermal transport property, the thermal conductivity of CNT-based composites is still low due to the poor interfacial thermal conductance (ITC) among CNTs.

Recently, thermal transport between CNTs has been investigated theoretically.[30,31] For overlapped CNTs, the intertube thermal conductance was enhanced by increasing the length and overlap area of CNTs. Even so, the interactions between CNTs are van der Waals (VDW) forces, thus thermal transport through non-bonded CNTs interface is not optimistic yet.

Here, we proposed a new SiC-CNT array composite structure as SiC/CNTs/SiC. As shown in Fig. 1(a), two CNT arrays with different diameters are respectively grown on each SiC surface, so that the two SiCs are bridged by overlapped CNTs. Besides, there are intertube carbon atoms with covalent bonding between two CNTs. This structure is feasible in practical experiments. The previous study has reported the joining of two single CNTs.[32]

In this paper, the ITC of SiC/CNTs/SiC was investigated by molecular dynamics simulations. We firstly gave a description of the model and simulation procedures. Secondly, the dependence of ITC on different numbers of intertube carbon atoms (N) was shown. Thirdly, the temperature dependence of ITC was also studied. Lastly, we performed theoretical analyses of the mechanism through atomic forces and vibrational density of states (VDOS).

**Simulation methods**

Classical non-equilibrium molecular dynamics (NEMD) method is employed to study the ITC of SiC/CNTs/SiC using large-scale atomic/molecular massively parallel simulator (LAMMPS) package.[33] The interatomic bonding interactions within SiC and CNTs are described by Tersoff potential[34,35] including both two-body and three-body potential terms, which has been widely used to study the thermal properties of SiC and CNT.[6,36,37] In addition, the non-bonded interactions between overlapped CNTs are described by the Lennard-Jones potential

$$V_{LJ}(r_{ij}) = 4\varepsilon[(\sigma/r_{ij})^{12} - (\sigma/r_{ij})^{6}] \quad (1)$$

where $\varepsilon$ is the depth of the potential well, $\sigma$ is the finite distance at which the interatomic potential is zero, $r_{ij}$ is the distance between atom $i$ and $j$. The Lennard-Jones parameters are $\varepsilon_{SC-PC} = 0.0028$ eV, $\sigma = 3.4$ Å, and the cutoff distance is set as 8.5 Å.

The simulation system consists of three parts (as shown in Fig. 1(a)), including the SiC of 4 × 4 × 4 unit cells at both ends and the overlapped CNTs in the middle. The diameters of inner-tube and outer-tube of CNTs are 0.678 nm and 1.356 nm respectively, corresponding to the chirality of (5, 5) and (10, 10). The atoms at boundaries in the longitudinal direction are fixed and the periodic boundary conditions are applied in the other two directions. The simulation procedures begin by equilibrating the system at 300 K with a Nose-Hoover thermostat for 50 ps. Then the hot region is raised to 320 K and the cold region is lowered to 280 K using Langevin thermostat for 2.5 ns. Lastly, it runs 5.0 ns to calculate the local average temperature and heat flow. The time step is set as 0.5 fs. SiC and CNTs are divided into several slabs to record the temperature profile. The heat flux is deduced by tallying the energy added to the "hot reservoir" and removed from the "cold reservoir". The ITC is calculated by the ratio of heat flux to the temperature difference.

$$G = J / \Delta T \qquad (2)$$

The intertube carbon atoms are randomly distributed along both axial and circumferential directions of the overlapped CNTs. We use a combination of time and ensemble sampling to obtain better average statistics. The result of each N represents an average of five independent simulations with different random distributions of intertube atoms.

Figure Fig. 1 shows a typical setup of the system and corresponding temperature profiles. The longitudinal view of the structure is shown in Fig. 1(a), and the corresponding cross section views are shown in the insets of Fig. 1(b-c). Thermal interfaces of the overlapped segment of CNTs and the whole part between SiC are concerned in this work. The results of temperature profiles show that the temperature differences

decrease obviously in the case of N = 2 compared with that in the case of N = 0, which indicates that the intertube atoms plays a positive role in enhancing the interfacial thermal transport.

**Results and Discussions**

Figure 2 illustrates the effects of the number of intertube carbon atoms on the ITC between two CNTs ($G_{CNTs}$) and the total ITC of the system ($G_{Total}$). $G_{CNTs}$ is smaller than the intertube thermal conductance of double-walled CNT in previous works,[31] because the length and overlap length of CNTs in this model are shorter. It is found that the ITC increase with the number of intertube atoms. The intertube atoms bridge the inner and outer CNTs through covalent bonds and form efficient channels for better heat transport, which lead to a sharp increase of ITC. Especially for N varying from 0 to 1, the ITC between two CNTs is enhanced by almost 10 times. It also indicates that the increasing rate of thermal conductance becomes lower with the increase of N. The ITC converges when the number of intertube atoms is beyond a critical value. Since more atoms are concentrated at the overlapped segment of CNTs, the additional intricate channels may constitute a jam which is not in favor of heat transport. Finally, compared with the case without intertube atom, $G_{CNTs}$ can be increased by almost two orders of magnitude, and $G_{Total}$ can be also improved by 20 times.

The enhancement of ITC by intertube atoms is a comprehensive result of two competing effects. The first effect is that, a high-efficiency channel and more phonon modes along radial direction are introduced by intertube atoms, which contribute to the heat transport, thus increase intertube thermal conductance. The second effect is that, the additional phonon modes and defects also open up new routes for phonon scattering, thus decrease the thermal conductivity of CNT and SiC.[38] In the case of SiC/CNTs/SiC, compared with SiC/CNTs interface interacted by covalent bonds, the main hindrance of thermal conductance comes from overlapped CNTs interacted by VDW forces. Moreover, compared with the effect of boundary scattering and mismatch of phonon modes between intratube high-frequency modes and intertube low-frequency modes, phonon scatterings induced by additional phonon modes are negligible in SiC/CNTs/SiC. Therefore, the first effect of more phonon modes introduced by intertube atoms is dominating and conducive to phonon transport, which leads to the increase of ITC. Furthermore, the fact that thermal conductance increases dramatically and then converges as more intertube atoms are added, also confirms that the first effect of more phonon modes is stronger than the secondary effect of phonon-phonon scattering.

The temperature dependences of ITC between two CNTs are shown in figure 3. A monotonic increase is observed in the ITC between two CNTs as a function of temperature. This is in striking contrast to the thermal conductivity behavior of CNTs and SiC, where the thermal conductivity at higher temperatures is found to be lower.[37,39-41] The monotonic increase of ITC is also observed in previous work.[28] It might be due to the fact that as the temperature increases, more phonons are excited in both SiC and CNT, and thus, contribute more to the ITC. In general, at high temperatures, the phonon-phonon scattering begins to dominate, thus hinders phonon transport in SiC and CNT. In the present case, however, the intertube thermal transport is considerably poorer than that in CNT and SiC, higher temperature excites more phonon modes along radial direction, giving rise to a better phonon-phonon coupling at the interface.

In order to reveal the differences of atomic interactions for cases with and without intertube atoms, Figure 4 shows the probability distribution functions of atomic forces along radial, axial and tangential directions, and vibrational density of states (VDOS) along radial direction of the atom at the connection of outer CNT. Figure 4(a) presents an extension of atomic forces along radial direction for the case with intertube atoms, which denotes stronger interactions and better thermal transport between CNTs. By contrast, there are slight changes of axial and tangential atomic forces, as shown in Fig. 4(b-c). VDOS is an effective analysis method to reflect the change of phonon vibrations. Figure 4(d) suggests that more phonon modes along radial direction are excited due to the intertube atoms. Compared with the case without intertube atom, the covalent bonding induced by intertube atoms result in modifications of phonon modes along radial direction and better thermal transport through interface.

**Conclusions**

We proposed a new feasible way to enhance the ITC of SiC through adding bonding atoms between overlapped CNTs. By non-equilibrium molecular dynamics simulations, the most important result is that the ITC is enhanced dramatically by adding intertube atoms, and converges with the increase of the number of intertube atoms. Compared with the case without intertube atom, $G_{CNTs}$ can be increased by almost two orders of magnitude, and $G_{Total}$ can be also improved by 20 times. The probability distributions of atomic forces and the vibrational density of states indicate that the covalent bonding induced by intertube carbon atoms strengthens intertube interactions and excites more phonon modes along radial direction, which leads to a better thermal transport. Our investigations may provide some guidance on enhancing the ITC of CNT-based composites.

**Acknowledgements**

The work was supported by the National Natural Science Foundation of China No. 51576076 (N. Y.) and No. 51576077 (X. H.), and the Self-Innovation Foundation of HUST No. 2015QN037 (C. D.). The authors acknowledge stimulating discussions with Quanwen Liao. The authors thank the National Supercomputing Center in Tianjin (NSCC-TJ) and the High Performance Computing Center Experimental Testbed in SCTS/CGCL for providing help in computations.

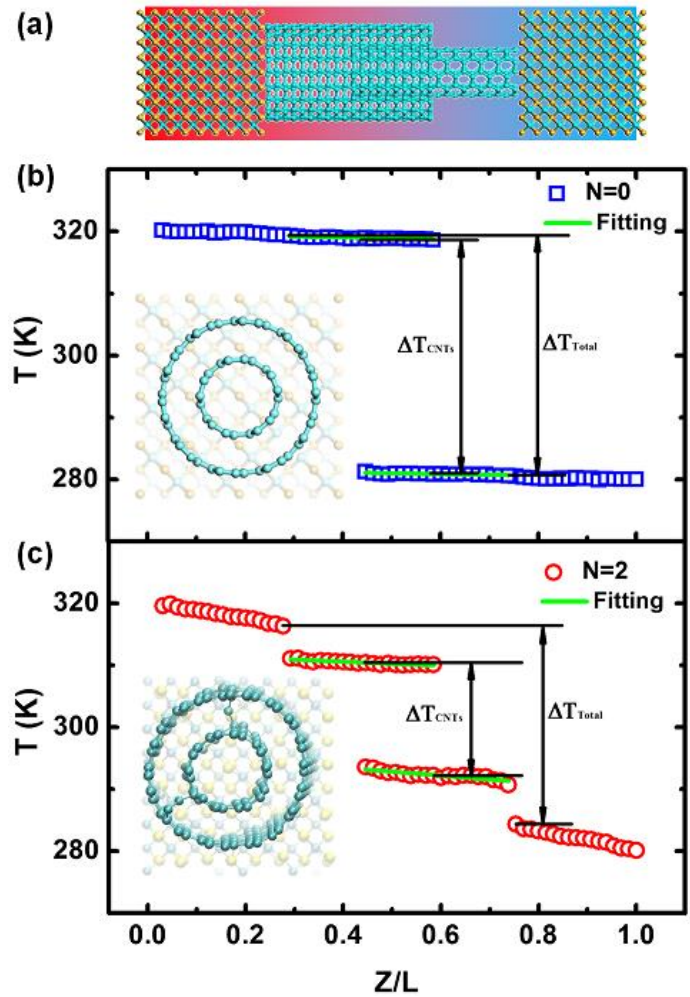

Figure 1. (a) Longitudinal view of simulation system, (b) and (c) temperature profiles and cross section views of simulation system for the cases of N = 0 and N = 2. The blue and yellow atoms represent carbon (C) and silicon (Si) respectively. N denotes the number of intertube atoms. The overlapped segment of CNTs and the whole parts between SiC are considered as the thermal interfaces of two CNTs and the whole simulation system respectively.

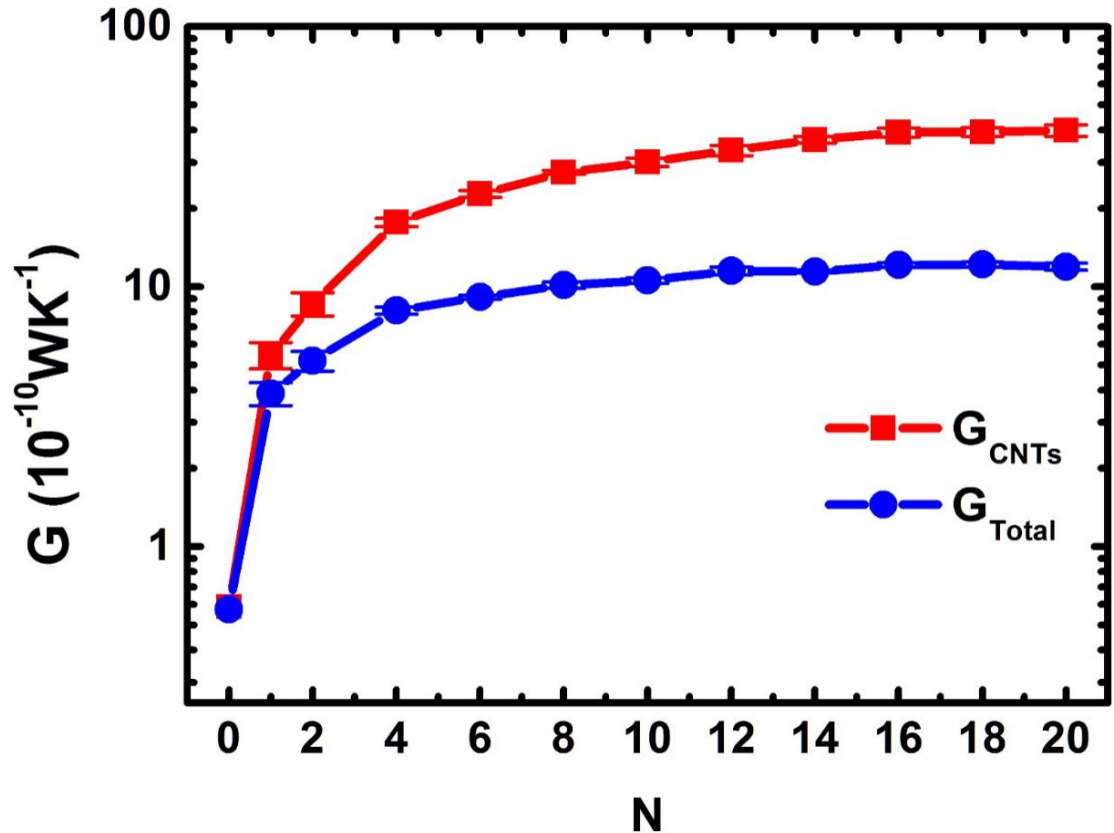

Figure 2. The interfacial thermal conductance (G) between two CNTs ($G_{CNTs}$) and the total thermal conductance ($G_{Total}$) of simulation system at room temperature as functions of the number of intertube atoms (N). The interfacial thermal conductance shows a sharp increase from N = 0 to N = 1. Both $G_{CNTs}$ and $G_{Total}$ converge gradually with the increase of N. Finally, $G_{CNTs}$ is enhanced by two orders of magnitude, and $G_{Total}$ is enhanced by almost 20 times as well.

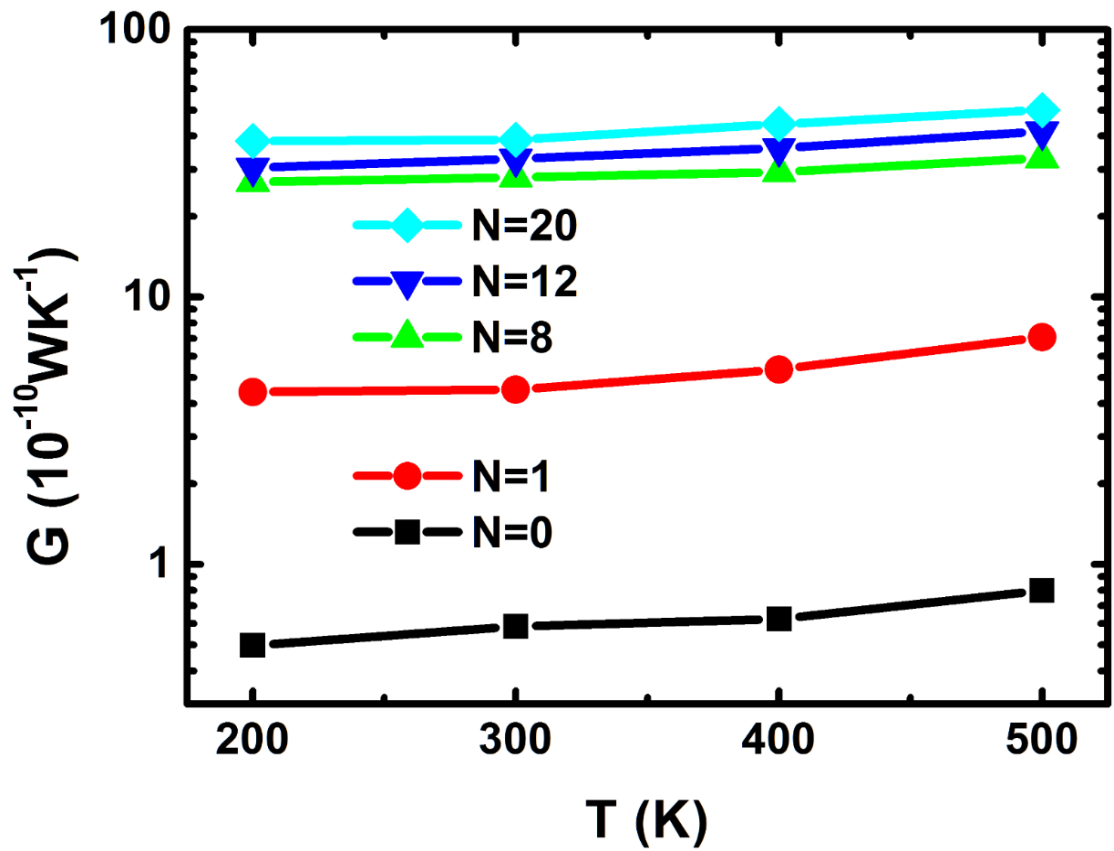

Figure 3. The temperature dependence of interfacial thermal conductance (G) between two CNTs for some typical cases of different N. Monotonic increases are observed as a function of temperature for all the cases.

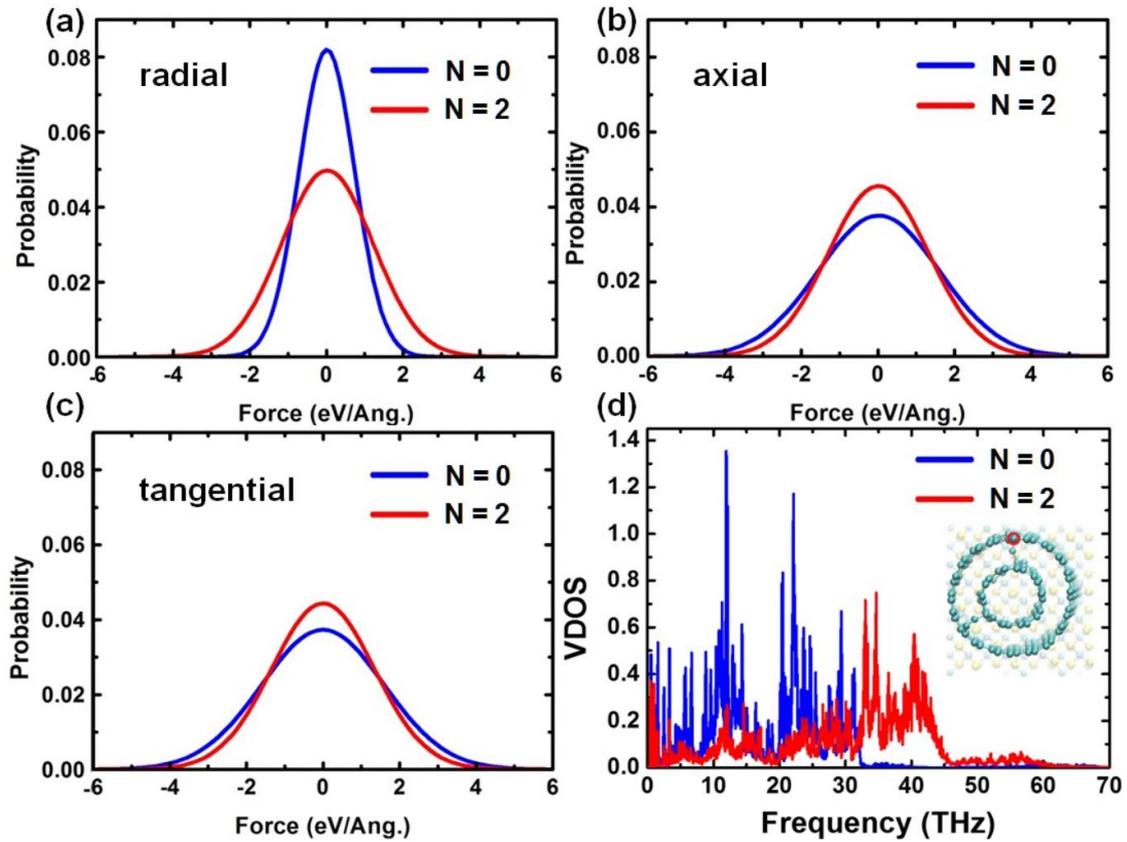

Figure 4. The probability distributions of atomic forces along (a) radial, (b) axial and (c) tangential directions, and (d) vibrational density of states (VDOS) along radial direction of the atom at the connection of outer CNT (red circle in (d)) for the cases of $N = 0$ and $N = 2$. Compared with the case without intertube atom, additional atoms lead to modifications of atomic forces and phonon modes. Covalent bonding induced by intertube atoms strengthens the interaction along radial direction, which is conducive to intertube thermal transport. And more phonon modes along radial direction are excited because of the strong covalent bonding, which also contribute to enhance interfacial thermal conductance.